\documentclass[iop,twocolumn]{aastex631}
\usepackage{amsmath}

\usepackage{gensymb}
\usepackage{hyperref}
\usepackage{graphicx,latexsym}

\newcommand\xone{x_1}
\newcommand\xtwo{x_2}
\newcommand\xthree{x_3}

\newcommand\Msun{\; {\rm M}_{\odot}}

\newcommand\kms{\; {\rm km/s}}
\newcommand\pc{\;{\rm pc}}

\newcommand\kpc{\;{\rm kpc}}

\newcommand\Gyr{\;{\rm Gyr}}

\newcommand\C{{\mathcal C}}

\newcommand\simgt{\lower.5ex\hbox{$\; \buildrel > \over \sim \;$}}
\newcommand\simlt{\lower.5ex\hbox{$\; \buildrel < \over \sim \;$}}
\newcommand{\RNum}[1]{\uppercase\expandafter{\romannumeral #1\relax}}

\newcommand\PSA{$\Omega_p-A_2$}
\newcommand\BS{$A_2$}
\newcommand\tbar{$\tau_\mathrm{bar}$}
\newcommand\PS{$\Omega_p$}
\newcommand\LzA{$\Delta L_z-A_2$}
\newcommand\Rratio{$\cal R$}
\newcommand\Rd{$R_d$}
\newcommand\RadiusRatio{$r_b/R_d$}
\newcommand\agama{{\sc agama}}

\def\spose#1{\hbox to 0pt{#1\hss}}
\def\dt{\spose{\raise 1.0ex\hbox{\hskip2pt$\mathchar"201$}}}

\defcitealias{Zheng2025}{Paper I}
\defcitealias{Zheng_taubar2025b}{Paper II}

\shorttitle{bulge effects on bar formation}
\shortauthors{Zheng et al.}

\begin{document}

\title{Comparison of Bar Formation Mechanisms. IIIA. The role of classical bulges in spontaneous bar formation}

\author[0000-0001-7707-5930]{Yirui Zheng}
% \correspondingauthor{Yirui Zheng}
% \email{yiruizheng@sjtu.edu.cn}
\affil{School of Physics and Astronomy, Anqing Normal University, Anqing 246011, P.R. China} 
\affil{Institute of Astronomy and Astrophysics, Anqing Normal University, Anqing 246133, P.R. China}
\affiliation{Department of Astronomy, School of Physics and Astronomy, Shanghai Jiao Tong University, 800 Dongchuan Road, Shanghai 200240, P.R. China, \href{mailto:jtshen@sjtu.edu.cn}{jtshen@sjtu.edu.cn}}
% \affiliation{Key Laboratory for Particle Astrophysics and Cosmology (MOE) / Shanghai Key Laboratory for Particle Physics and Cosmology, Shanghai 200240, P.R. China}
% \affiliation{National Key Laboratory of Dark Matter Physics / Shanghai Key Laboratory for Particle Physics and Cosmology, Shanghai 200240, P.R. China}
\affiliation{State Key Laboratory of Dark Matter Physics, School of Physics and Astronomy, Shanghai Jiao Tong University, Shanghai 200240, P.R. China}
\affiliation{Key Laboratory for Particle Astrophysics and Cosmology (MOE) / Shanghai Key Laboratory for Particle Physics and Cosmology, Shanghai 200240, P.R. China}

\author[0000-0001-5604-1643]{Juntai Shen}
\correspondingauthor{Juntai Shen}
\email{jtshen@sjtu.edu.cn}
\affiliation{Department of Astronomy, School of Physics and Astronomy, Shanghai Jiao Tong University, 800 Dongchuan Road, Shanghai 200240, P.R. China, \href{mailto:jtshen@sjtu.edu.cn}{jtshen@sjtu.edu.cn}}
\affiliation{State Key Laboratory of Dark Matter Physics, School of Physics and Astronomy, Shanghai Jiao Tong University, Shanghai 200240, P.R. China}
\affiliation{Key Laboratory for Particle Astrophysics and Cosmology (MOE) / Shanghai Key Laboratory for Particle Physics and Cosmology, Shanghai 200240, P.R. China}

% \author[0000-0002-1378-8082]{Xufen Wu}
% \affiliation{Department of Astronomy,
% University of Science and Technology of China, Hefei 230026, P.R. China}
% \affiliation{School of Astronomy and Space Science, University of Science and Technology of China, Hefei 230026, P.R. China}

\author[0000-0001-8962-663X]{Bin-Hui Chen}
\affiliation{Tsung-Dao Lee Institute, Shanghai Jiao Tong University, Shanghai 200240, P.R. China}
\affiliation{Department of Astronomy, School of Physics and Astronomy, Shanghai Jiao Tong University, 800 Dongchuan Road, Shanghai 200240, P.R. China, \href{mailto:jtshen@sjtu.edu.cn}{jtshen@sjtu.edu.cn}}
% \affiliation{Key Laboratory for Particle Astrophysics and Cosmology (MOE) / Shanghai Key Laboratory for Particle Physics and Cosmology, Shanghai 200240, P.R. China}
% \affiliation{National Key Laboratory of Dark Matter Physics / Shanghai Key Laboratory for Particle Physics and Cosmology, Shanghai 200240, P.R. China}
\affiliation{State Key Laboratory of Dark Matter Physics, School of Physics and Astronomy, Shanghai Jiao Tong University, Shanghai 200240, P.R. China}

%===============================================================================

\begin{abstract}

We run a suite of $N$-body simulations to investigate how classical bulges affect bar formation and properties under the internal formation mechanism. 
% We add bulges of varying mass and compactness to disk galaxy models that are susceptible to bar instability.
% The models are simulated in isolation, and we examine the resulting bar pattern speeds and growth timescales. 
We incorporate bulges of varying mass and compactness into disk galaxy models and evolve them in isolation to examine the resulting bar pattern speeds and growth timescales.
A more massive/compact bulge increases the Toomre $Q$ stability parameter and the circular velocity in the central region, while decreasing the disk mass fraction. 
It therefore delays the onset of bar formation and increases the bar growth timescale; sufficiently strong bulges can suppress bar formation entirely. 
During the formation stage, bars exhibit higher initial pattern speeds and faster deceleration rates when the bulges become more massive or compact.
This faster deceleration persists after the bar buckling phase, leading to slower-rotating bars in the secular growth stage.  
However, when the bulge's ``diluting'' effect on the measured bar strength is removed or reduced, all bars within the same disk share similar distributions in the pattern speed--bar strength ($\Omega_p$--$A_2$) space during the secular growth stage. 
They also show comparable ratios of the co-rotation radius to the bar length ($\mathcal{R}=R_{\mathrm{CR}}/R_{\mathrm {bar}}$) in this stage.
These results suggest that the bulge's influence on the pattern speed is more significant during the bar formation stage, while in the secular growth stage, the bulge's effect may be less important, and the disk component dominates the pattern speed evolution.

\end{abstract}


%-----------
%-- Sect. 1
%-----------

\section{Introduction}
\label{sec:intro}

% standard begin
% bulge effect
%% pattern speed
%% growth timescale
% external mechanism
% paper I and II

Galactic bars are common and important structures of spiral galaxies. 
In the local universe, optical surveys find bars in nearly 50\% of disk galaxies, a fraction that rises to approximately two-thirds in infrared bands\citep{Marinova2007, MenendezDelmestre2007, Erwin2018, Lee2019bar}.
Although less frequent at higher redshifts, bars are still present, constituting roughly 13\% of galaxies at $2<z\leq3$ \citep{LeConte2024} and have been detected as early as $z\sim4$ \citep{Guo2024}. 
These structures play an important role in driving the evolution of their host galaxies by fueling gas inflows, regulating star formation, and contributing to the formation of pseudo-bulges \citep{masters2011galaxy, Li2015, Lin2017, Lin2020, Iles2022}.
Given their ubiquity and influence, understanding the formation and evolution of bars is critical in galactic dynamics.

Galactic bars can form via two primary mechanisms: internal instability and external perturbations. 
The internal, or spontaneous, mechanism arises from gravitational instability within the disk itself \citep{hohl1971numerical, ostriker1973numerical, sellwood2014secular, Lokas2019iso}.
A well-established mechanism for spontaneous bar formation is the swing amplification feedback loop that was first introduced by \citet{Goldreich1965} and \citet{Julian1966}, later elaborated in detail by \citet{Toomre1981}, and more recently revisited and clarified by \citet{Binney2020}.
However, the presence of a strong bulge or central mass concentration can introduce inner Lindblad resonances (ILR), which disrupts this feedback loop and thereby suppresses bar formation \citep{Toomre1981, Binney2008}. 
If not disrupted, the feedback loop predicts an exponential growth in bar strength.

The growth rate, or equivalently, the formation timescale of bars formed via internal instability, is found to be influenced by several factors, including the bulge-to-disk mass ratio \citep{Kataria2018}, the Toomre $Q$ parameter \citep{Hozumi2022, Worrakitpoonpon2025}, and the disk thickness \citep{Ghosh2023}. \citet{Fujii2018} reported an exponential decrease in formation timescale with increasing disk mass fraction—the ``Fujii relation''—later confirmed by \citet{BlandHawthorn2023}. 
The latter study introduced a key methodological improvement: it quantified the exponential growth timescale predicted by the swing amplification loop to measure bar formation instead of using the time the bar strength exceeds a fixed threshold (e.g., $A_{2, max}>0.2$).
\citet{Chen2025} extended the Fujii relation in disk galaxies to a three-dimensional parameter space of disk mass fraction, Toomre $Q$, and disk thickness.
Their empirical relation revealed a linear dependence of the bar
growth timescale on both $Q$ and the disk thickness.

The disk mass fraction is established as the key factor governing the bar formation timescale \citep{Fujii2018, BlandHawthorn2023, BlandHawthorn2024, BlandHawthorn2025, Chen2025}. Since the presence of a bulge reduces this fraction, the bulge component is consequently expected to have a significant effect on the timescale for bar formation.
In addition, bulges increase the Toomre $Q$ parameter, particularly in the inner disk, and may also influence the disk thickness. Therefore, bulges can affect bar formation through these two additional factors as well.

Several studies have examined how bulges or central mass concentrations (CMCs) suppress bar formation, although a universal, quantitative criterion for suppression remains elusive due to the problem's inherent complexity \citep[e.g.,][]{Kataria2018, Saha2018, Jang2023, Worrakitpoonpon2025}. 
Nonetheless, consistent trends regarding the bulge's influence on bar properties have emerged.
\citet{Kataria2018} demonstrated that bars have higher pattern speeds in galaxies with more massive bulges, and that the pattern speed decelerates more rapidly as the bulge mass fraction increases. 
\citet{Jang2023} confirmed that bars tend to be stronger, longer, and rotate more slowly in galaxies with less massive and less compact bulges. 
This trend is further supported by \citet{McClure2025}, who also validated it observationally in a sample of 210 MaNGA disk galaxies, finding that slow bars are preferentially associated with weaker bulges.

% Worrakitpoonpon2024
% First, disk stabilization requires both high Q and CMC. Either high Q or high CMC only results in slow bar formation.

% Jang2023
% Bars tend to be stronger and longer, and to rotate more slowly, in galaxies with a less massive and less compact bulge and halo. All bars formed in our models correspond to slow bars.

% McClure, Rachel Lee25
% We compare these results with a sample of 210 MaNGA disk galaxies, finding that slow bars--indicative of older systems--are preferentially associated with weaker bulges.

% Sandeep 
% pattern speed, deceleration

Besides internal instabilities, bars can also be triggered by external perturbations, such as tidal forces from galaxy clusters, flyby encounters, collisions, or mergers \citep{byrd1986tidal, gauthier2006substructure, Lokas2016, MartinezValpuesta2017, Lokas2019tidal}. 
In this series of papers, we aim to comprehensively compare the properties of bars formed via internal instability versus those formed under external perturbations.

In the first paper of this series \citep[][Paper~I hereafter]{Zheng2025}, we conducted a suite of controlled $N$-body simulations in which bars form either internally or under tidal perturbations. We constructed three pure disk galaxy models of varying dynamical ``hotness'' and simulated their evolution both in isolation and under flyby interaction. 
We then measured bar properties, comparing pattern speeds in Paper~I and extended the comparison to exponential growth rates in the second paper \citep[][Paper~II hereafter]{Zheng_taubar2025b}.
For cold and warm disks—which can form bars spontaneously in isolation—tidal interactions promote bars to a more advanced evolutionary stage. 
Nevertheless, these tidally induced bars exhibit pattern speeds and growth timescales similar to their spontaneously formed counterparts within the same disk. 
In hot disks, which are stable against internal bar instability, a bar forms only under external perturbations and rotates more slowly than bars in cold or warm disks. 
Therefore, if ``tidally induced bars'' refer exclusively to those in otherwise bar-stable galaxies, they do rotate more slowly. 
However, this difference is attributed to the intrinsic properties of the host galaxy rather than to the formation mechanism itself.
As for growth timescales of these bars, their growth deviates from an exponential profile and is better described by a linear function, 
which may suggest that these tidally forced bars do not adhere to the swing amplification mechanism.

The third objective of this paper series is to investigate how classical bulges influence bar formation and properties. 
In the present work (Paper~IIIA), we focus on their effects under the \textit{internal} formation mechanism. 
The upcoming companion study (Paper~IIIB) will extend this investigation to the \textit{external} formation scenario, examining whether the influence of bulges differs between bars formed via these two distinct mechanisms.

Paper IIIA is organized as follows. 
In Section~\ref{sec:sims}, we describe the construction of
galaxy models with varying bulge properties.
Section~\ref{sec:patt_sp} investigates the influence of
classical bulges on the pattern speed of spontaneously formed
bars. 
In Section~\ref{sec:growth_timescale}, we perform an
exponential fit to the bar growth to quantify how bulge
properties affect the bar growth timescale and onset time.
Finally, we summarize our findings in Section~\ref{sec:summary}.

%-----------
%-- Sect. 2
%-----------

\section{Simulations}
\label{sec:sims}

\subsection{Galaxy models}
\label{subsec:gal_models}

\begin{figure*}
  \centering
  \includegraphics[width=\textwidth]{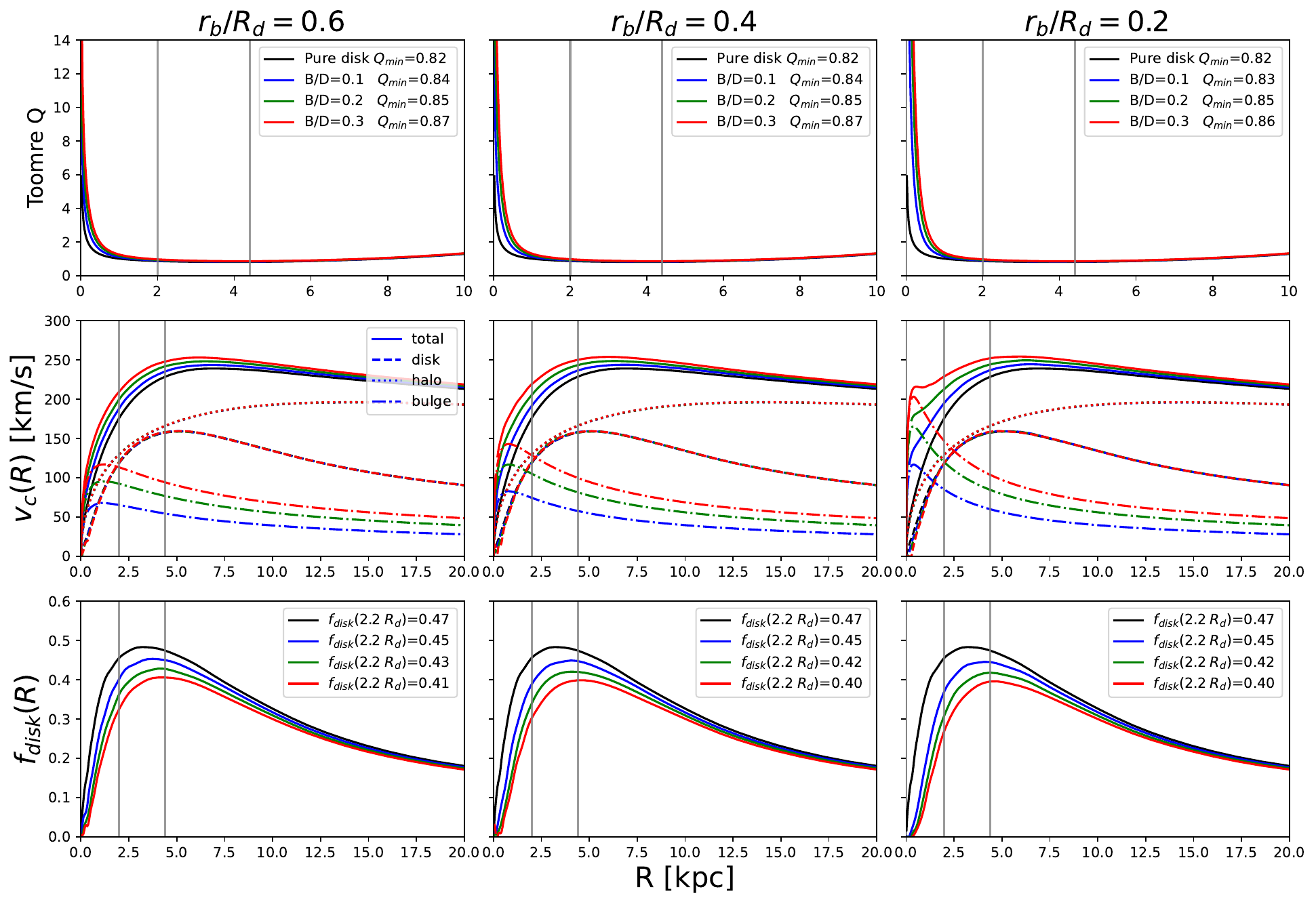}
  % \caption{test}
  \caption{Toomre $Q$ profiles (top row), circular speeds (middle row), and disk mass fraction $f_{\text{disk}}$ (bottom row) for the cold series models.
  Columns show increasing bulge compactness (left to right), with the scale radius (length) ratio \RadiusRatio$=0.6$, 0.4, and 0.2.  
  Solid lines trace models of varying bulge-to-disk mass ratio ($B/D$), with the minimum $Q$ value and $f_{\text{disk}} (2.2\;R_d)$ noted in the legend. 
  The middle row decomposes $v_c(R)$ into contributions from bulge (dash-dotted), stellar disk (dashed), and DM halo (dotted) components. Vertical gray lines indicate 1$R_d$ and 2.2$R_d$ (where $R_d=2$\kpc).
  }
\label{fig:model}
\end{figure*}

In \citetalias{Zheng2025} and \citetalias{Zheng_taubar2025b}, we employ a suite of pure-disk models generated with the \agama\ software \citep{AGAMA2019}. 
These models share a common density profile but differ only in the velocity dispersion of their stellar disks. They comprise two components: an exponential, quasi-isothermal stellar disk and a truncated Hernquist dark matter (DM) halo. 
The stellar disk has a mass of $M_* = 3.6 \times 10^{10} \Msun$, a scale length of $R_d = 2 \kpc$, and a scale height of $h_z = 0.4 \kpc$. 
The DM halo is characterized a mass of $M_{\rm halo} = 3.6 \times 10^{11} \Msun$ and a scale radius of $a = 13.7 \kpc$. 
For comprehensive details on both components, we refer the reader to \citetalias{Zheng2025}.

In this study, we further incorporate a classical bulge component into the galaxy model. 
Its density distribution is given by a truncated Hernquist profile \citep{hernquist1990analytical}:
\begin{equation}
  \rho_{\rm b} = \frac{M_{\rm b}}{2\pi} \frac{r_b}{r(r+r_b)^3} \times \exp \left[-(r_b/r_{\rm cut,b})^2 \right].
\end{equation}
Here, the density profile consists of the standard Hernquist form—parameterized by the total mass $M_{\rm b}$ and the scale radius $r_b$—multiplied by an exponential cutoff term. This cutoff, governed by the cutoff radius $r_{\rm cut,b}$, acts to suppress the density at large radii. In all models, we set $r_{\rm cut,b} = 8\;r_b$.

To create models with varying bulge prominence, we consider a range of bulge-to-disk mass ratios: $B/D=0.1$, 0.2, and 0.3.  This yields bulge masses of $M_{\rm b} = 3.6 \times 10^9 \Msun$, $7.2 \times 10^9 \Msun$, and $1.08 \times 10^{10} \Msun$, respectively. 
Similarly, we explore different bulge compactness by varying the ratio of the bulge scale radius to the disk scale length, \RadiusRatio$=0.2$, 0.4, 0.6, with smaller values indicating a more compact bulge. The corresponding bulge scale radii are $r_b = 0.4 \kpc$, $0.8 \kpc$, and $1.2 \kpc$, respectively.

The stellar disk velocity distribution in \agama\ is set by an action-based distribution function (DF).
The key stability parameter from the DF is the radial velocity dispersion, which declines exponentially with a scale length $R_{\sigma, R} = 2\;R_d$. 
We define three model series—cold, warm, and hot disks—by setting the central radial velocity dispersion to $\sigma_{R,0} = 73 \kms$, $124 \kms$, and $226 \kms$, respectively. 
Each series comprises one pure disk model and nine galaxy models with bulges (varying in mass and compactness) as described previously.

We quantify the stability of the stellar disk by the Toomre $Q$ parameter, which is defined as:
\begin{equation}
    Q = \frac{\kappa \sigma_R}{3.36 G \Sigma},
\label{eq:toomreQ}
\end{equation}
with $\kappa$, $\sigma_R$, $G$, and $\Sigma$ representing the
epicycle frequency, radial velocity dispersion, gravitational
constant, and disk surface density, respectively. 
The top row of \autoref{fig:model} shows the $Q$ profiles for the cold model series.  
Relative to the pure disk case, $Q$ is similar in the outer disk but exhibits a substantial increase in the inner region. This increase is particularly pronounced for more massive (larger $B/D$) and more compact (smaller \RadiusRatio) bulges.

The middle row of \autoref{fig:model} presents the circular speed curves $v_c(R)$ for the cold series, along with decompositions into contributions from the halo, stellar disk, and bulge components.
% Contributions from the halo, stellar disk, and bulge are indicated by dotted, dashed, and dash-dotted lines, respectively.
The circular speed increases with bulge mass, particularly in the inner region. Furthermore, the inner profile of $v_c(R)$ is significantly modified by the presence of a compact bulge.

Following \citet{Fujii2018} and \citet{BlandHawthorn2023}, we compute the $f_{\text{disk}}$ parameter as the mass ratio of the disk to the total galaxy: 
\begin{equation}
f_{\text{disk}}(R) =  \frac{v^2_{c,\text{disk}}(R)}{v^2_{c,
\text{total}}(R)}.
\end{equation}
This parameter is shown in the bottom row of \autoref{fig:model}; the legends indicate its value at $R=2.2\;R_d$, the radius where the rotation curve typically peaks, and which is used in the Fujii relation.
The bulge component reduces $f_{\text{disk}}$ as anticipated, with more massive bulges producing lower values.
% In contrast, the compactness of the bulge only affects $f_{\text{disk}}$  in the inner region but has limited influence on  $f_{\text{disk}}(2.2\;R_d)$.
In contrast, bulge compactness significantly affects $f_{\text{disk}}$ only in the innermost region, whereas it exerts limited influence on $f_{\text{disk}}$ at $2.2\;R_d$.

The results for the warm and hot model series follow the same trend and are therefore omitted for brevity.

\subsection{Simulation details}
\label{subsec:simdetails}

To study the influence of the bulge on spontaneous bar formation, each galaxy model is simulated in isolation for 6\Gyr\ with \texttt{GADGET-4} \citep{Springel2005, Springel2021}. 
The particle numbers are as follows: one million for the DM halo, $0.5$ million for the stellar disk, and $0.5\times (B/D)$ million for the bulge.
This corresponds to particle masses of $3.6 \times 10^5 \Msun$ for the halo and $7.2 \times 10^4 \Msun$ for the stellar components. 
The gravitational softening lengths are 57\pc\ for DM particles and 23\pc\ for stars.
% The gravitational softening lengths are 23\pc\ for stars and 57\pc\ for DM particles.

Although the mass resolution in our simulations is moderate, it is adequate for tracking bar growth. 
This is supported by the findings of \cite{Dubinski2009}, who demonstrated convergence in key aspects of bar evolution (such as growth, pattern speed, dark matter halo density profile, and nonlinear orbital resonance analysis) when comparing simulations with one million to ten million halo particles.

% -----------
% -- Sect. 3
% -----------

\section{Bulge effects on bar pattern speed}
\label{sec:patt_sp}

\subsection{Bar strength and pattern speed}
\label{subsec:barpro}

\begin{figure*}
  \centering
  \includegraphics[width=\textwidth]{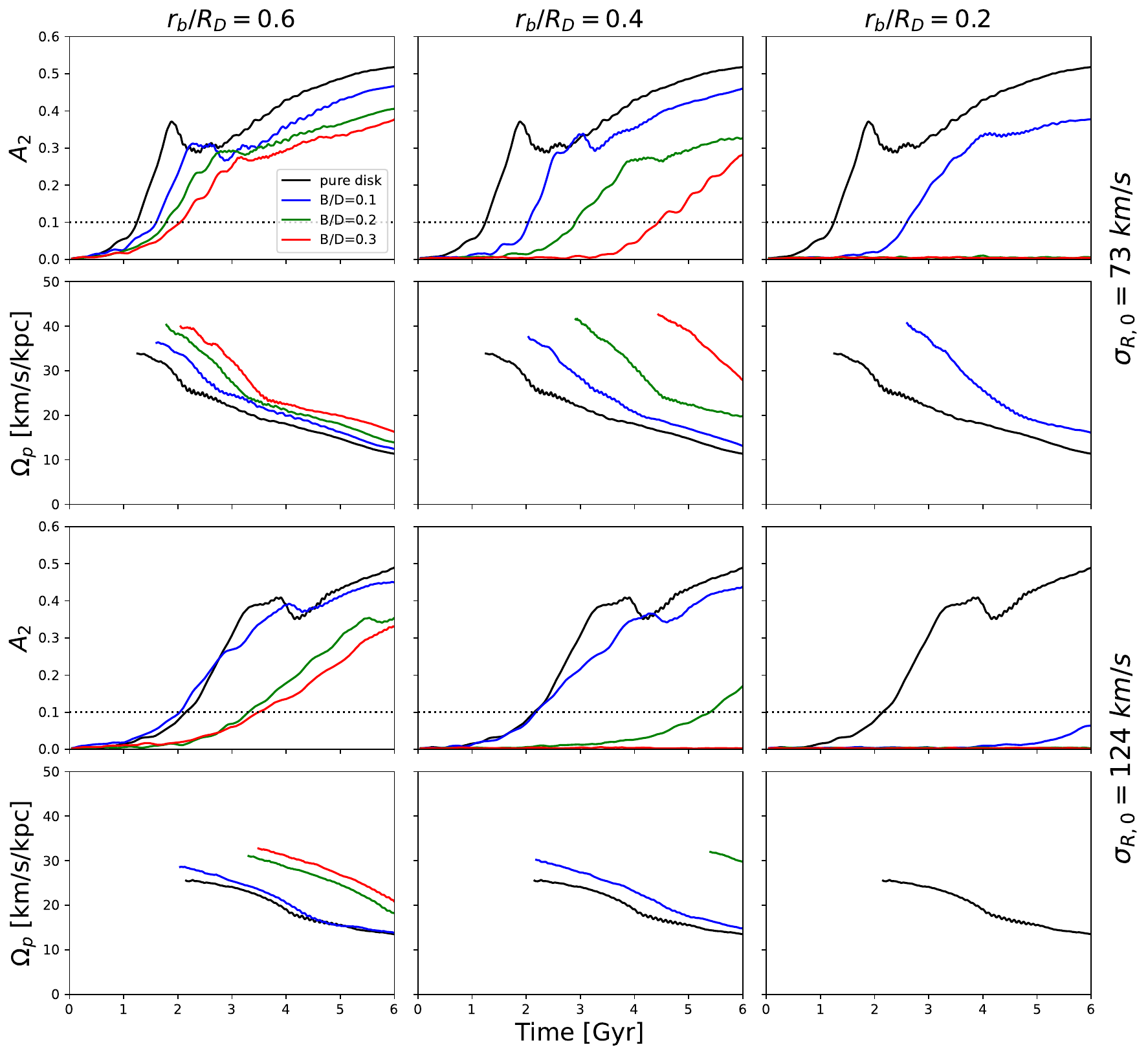}
  % \caption{test}
  \caption{Bar strength $A_2$ (odd rows) and pattern speed \PS\ (even rows) for different models. 
  Models are arranged by increasing bulge compactness from left to right, with \RadiusRatio$=0.6$, 0.4, and 0.2.
  Rows show results of different dynamical hotness: the cold series ($\sigma_{R,0}=73 \kms$, top two rows) and the warm series ($\sigma_{R,0}=124 \kms$, bottom two rows). The results for the hot series ($\sigma_{R,0}=226 \kms$) are not shown as no bar forms in isolation for any of the models in this series.
  }
\label{fig:bar_pro}
\end{figure*}

Bar strength \BS\ and pattern speed \PS\ are quantified via
Fourier analysis.
We compute the $m=2$ Fourier coefficients within a cylindrical radius $R \leq 4R_d$ (i.e., $R \leq 8\;\kpc$ for our models):
\begin{subequations}
\label{eq:a2b2}
\begin{align}
    a_2 &= \frac{ \sum_{i=1}^{N} m_i \cos(2\phi_i)} {\sum_{i=1}^{N} m_i}, \\
    b_2 &= \frac{ \sum_{i=1}^{N} m_i \sin(2\phi_i)} {\sum_{i=1}^{N} m_i},
\end{align}
\end{subequations}
summing over particles with mass $m_i$ and azimuth angle $\phi_i$.
% The limit of $4\;R_d$ is selected as it approximately corresponds to the maximum radius of the bar in our simulations.
The bar strength $A_2$ is then decided as:
\begin{equation}
    A_2 = \sqrt{a_2^2 + b_2^2},
    \label{eq:bar_strength}
\end{equation}
while the position angle of the bar $\phi_{\mathrm {bar}}$ is :
\begin{equation}
   \phi_{\mathrm {bar}} = \frac{1}{2} \arctan({b_2}/{a_2}).
\end{equation}
The pattern speed \PS\ is defined as the time derivative of the position angle, i.e.,
\begin{equation}
    \Omega_p = {\Delta \phi_{\mathrm {bar}}}/{\Delta t}.
\end{equation}
We measure $\Omega_p$ only when $A_2 \geq 0.1$, since
$\phi_{\mathrm{bar}}$ is not well-defined during the early stages of bar formation.

\autoref{fig:bar_pro} shows the evolution of bar strength $A_2$ and pattern speed \PS\ for our models during the isolated simulations.  
The raw \BS\ and \PS\ data may fluctuate due to transient features such as spurs or spiral arms attached at the ends of the bars.  
We therefore smooth both quantities with a 0.2\Gyr\ moving-average window to mitigate these fluctuations.

Results for the cold model series ($\sigma_{R,0}=73\kms$) are displayed in the top two rows. A bulge component delays bar formation compared to the pure disk model, which will be further quantified in Section~\ref{sec:growth_timescale}.
At a given bulge mass ($B/D$), greater compactness (smaller \RadiusRatio) results in a longer bar formation delay. These trends align with the increased disk stability (higher Toomre $Q$) in the inner region caused by a stronger bulge, as seen in \autoref{fig:model}.
Moreover, for sufficiently massive ($B/D \geq 0.2$) and compact (\RadiusRatio$=0.2$) bulges, bar formation is completely suppressed over the full 6\Gyr\ simulation.
The suppression is physically anticipated. 
As demonstrated in \autoref{fig:model}, a strong bulge elevates the central circular speed and thereby the angular frequency. 
This may establish an inner Lindblad resonance (ILR), which prevents density waves from propagating and thus breaks the swing amplification feedback loop.
%  essential for bar growth. 
Consequently, bar formation is suppressed. 

We find that bars formed in models with more massive or compact bulges have higher pattern speeds at birth. 
This is consistent with the higher circular speed (and thus higher angular speed) in the inner regions of such models, as seen in \autoref{fig:model}.
Our finding is in agreement with several previous studies \citep{Kataria2018, Jang2023, McClure2025}, which similarly observed faster pattern speeds for bars in galaxies with more prominent bulges.

After their formation, bars undergo gradual deceleration—a well-established result commonly seen in the literature \citep{Weinberg1985Apr, Debattista2000, Athanassoula2003} and in our own previous work (\citetalias{Zheng2025}). 
This deceleration is driven by the capture of disk stars onto the outer parts of the bar and by the dynamical friction between the bar and the dark matter halo.
Bars formed in models with a more massive or compact bulge decelerate faster after formation, as indicated by the steeper gradient in their \PS\ evolution. This finding is consistent with \citet{Kataria2018}. 
The differences in bar pattern speed and deceleration rate are more clearly identified in the pattern speed--bar strength (\PSA) space. 
This phase-space representation accounts for the different evolutionary stages of bars and will be explored in Section~\ref{subsec:psa}.

The bottom two rows of \autoref{fig:bar_pro} show results for the warm model series ($\sigma_{R,0} = 124 \kms$). The presence of a bulge component similarly delays bar formation and raises the pattern speed at formation.  
More models in this series fail to form a bar within 6\Gyr;  the warmer disk is intrinsically less susceptible to the bar instability, and the bulge provides additional stabilization.
For high bulge compactness (\RadiusRatio$=0.2$), even a low-mass bulge ($B/D=0.1$) suffices to suppress the formation of identifiable bars ($A_2\ge0.1$) within 6\Gyr.

We omit the results for the hot model series, as none of these models form a spontaneous bar within 6\Gyr. The hot stellar disk is intrinsically stable against bar formation, as demonstrated in \citetalias{Zheng2025}, and the addition of a bulge further stabilizes the system. Consequently, the hot series models are excluded from the subsequent analysis in this paper.

\vspace{5mm}

\subsection{Pattern speed--bar strength (\PSA) space}
\label{subsec:psa}

\begin{figure*}
  \centering
  \includegraphics[width=\textwidth]{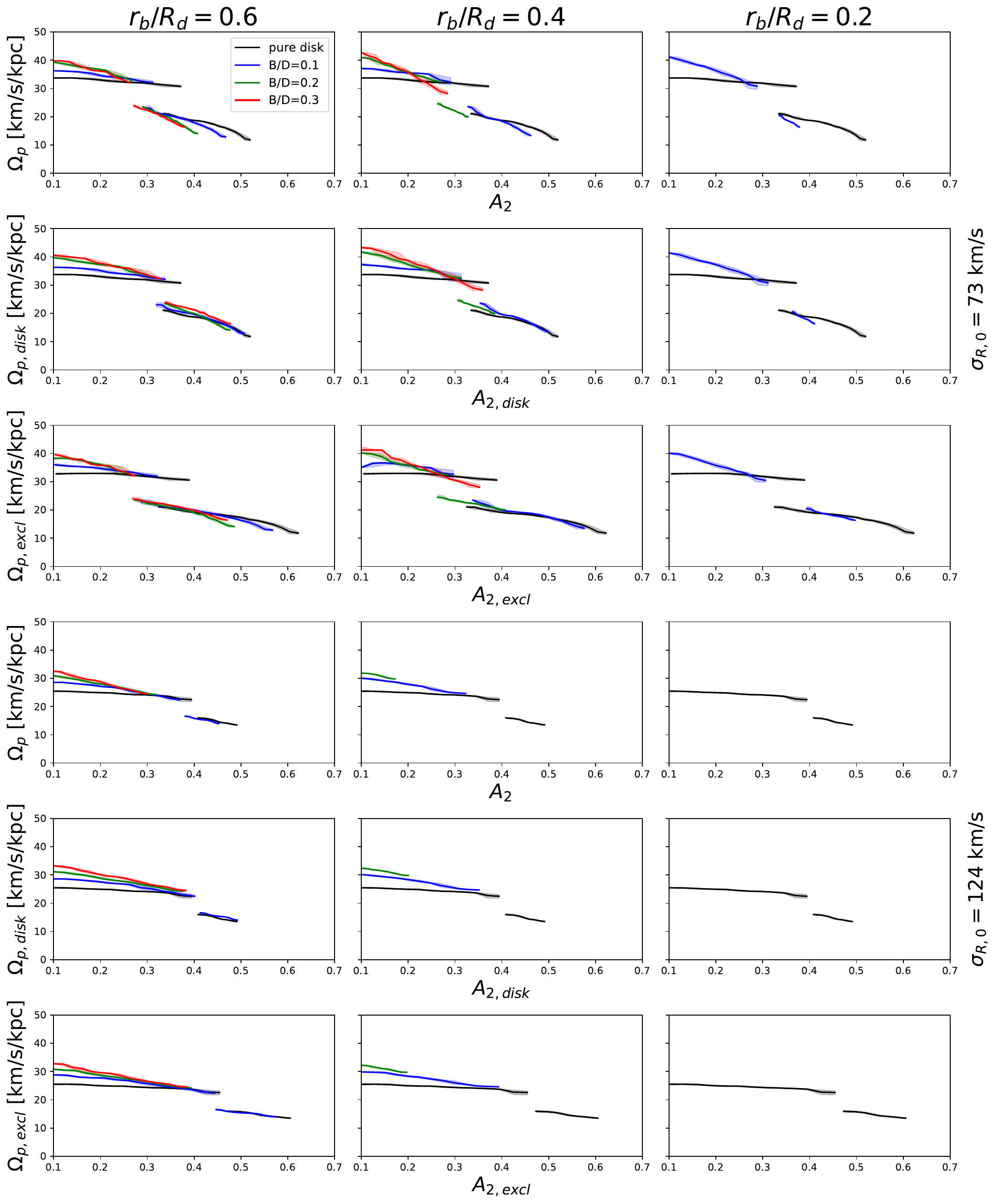}
  % \caption{test}
  \caption{
    The \PSA\ space of all simulations. Panel layout and color scheme follow \autoref{fig:bar_pro}. 
    Three definitions of bar strength $A_2$ are employed: 
    (1) \BS\ from all stars within $4\;R_d$ (rows 1 \& 4); 
    (2) $A_{2,\text{disk}}$ from disk particles only within $4\;R_d$, removing the bulge (rows 2 \& 5); 
    (3) $A_{2,\text{excl}}$ from all stellar particles within $4\;R_d$ but excluding the central $1\;R_d$, mitigating bulge influence (rows 3 \& 6). 
    In the bar formation phase, higher bulge mass leads to higher $\Omega_p$ at fixed bar strength. In the secular growth stage, models converge to similar $\Omega_p$ at fixed $A_2$ when the bulge contribution is excluded or reduced.
    }
\label{fig:PS_A2}
\end{figure*}

In \citetalias{Zheng2025}, we found that tidally induced bars in the cold and warm disks can appear to have lower pattern speeds at a given simulation time. 
However, they actually rotate as fast as their spontaneous counterparts in the same galaxy when bar strength is considered, as evidenced by their similar distributions in the pattern speed--bar strength (\PSA) space. 
This highlights that the different evolutionary stages of a bar—given its deceleration—can complicate a direct comparison of pattern speeds at fixed times. 
For the same reason, we compare the \PSA\ space across models with different bulge properties. 
This approach allows us to understand how the bulge influences the pattern speed at a fixed bar strength, thereby accounting for the fact that the bulge delays the bar's evolutionary stage.

The procedure for tracking bar evolution in the \PSA\ space is illustrated in Section~3.2 of \citetalias{Zheng2025}; see their Figure~5 for a representative example. 
In short, we select the \BS\ and \PS\ data from both the bar formation and secular growth stages. Within each stage, the data are binned by $A_2$, and we compute the median along with the 16th and 84th percentiles of \PS\ in each bin.
A key modification from \citetalias{Zheng2025} is that we smooth the \BS\ and \PS\ data with a moving average before binning, further mitigating the influence of transient fluctuations on the derived percentiles.

Distributions in the \PSA\ space for the cold series models are shown in the top row of \autoref{fig:PS_A2}. 
In the bar formation phase, models with more massive or compact bulges produce bars that start with a higher initial \PS\ and decelerate more rapidly as $A_2$ increases, reproducing the trends seen in \autoref{fig:bar_pro}. 
Since this comparison is made at fixed bar strength, it demonstrates that the differences in \PS\ are intrinsic properties linked to the bulge, not merely a consequence of the different evolutionary stages of bars at a given simulation time.
Interestingly, all bars converge to a comparable pattern speed by the end of the formation stage, despite attaining different maximum bar strengths. During the secular growth stage, bars in models with a stronger bulge continue to decelerate faster.
Consequently, at a given bar strength, they have a lower \PS\ than bars in models with a less massive or less compact bulge.

A reasonable concern is that the bulge's contribution could ``dilute'' the measured bar strength, especially during the secular stage when the bar is near saturation. 
Our Fourier analysis shows that the classical bulge saturates at a maximum $m=2$ amplitude of ${\sim}0.10$--$0.15$, significantly lower than the typical bar strength in our simulations.
Therefore, according to \autoref{eq:a2b2} and \ref{eq:bar_strength}, a more dominant bulge would indeed yield a lower \BS\ when the stellar disk has a similar level of distortion.
To assess whether bulge-induced dilution of \BS\ artificially steepens the $\Omega_p$ gradient, we recompute the bar strength and pattern speed using only disk particles. 
These quantities, labeled $A_{2,\text{disk}}$ and $\Omega_{p,\text{disk}}$, are used to regenerate the \PSA\ space, shown in the second row of \autoref{fig:PS_A2}.

Even with disk-only measurements, a more massive/compact bulge still leads to a higher initial $\Omega_p$ and faster deceleration at fixed $A_2$ during the bar formation stage. 
Strikingly, during the secular growth stage, the pattern speeds across models converge to similar values at fixed $A_2$, which is not previously observed. 
These results imply that the bulge exerts a stronger influence on the pattern speed during bar formation, whereas in the secular stage, the evolution of $\Omega_p$ is largely governed by the disk component, with the direct influence of the bulge possibly becoming less dominant.

In real observations, the clean separation of the disk and bulge components could be challenging. 
Therefore, we introduce an alternative, more practical method to reduce the bulge's effect on the measured bar strength. 
We simply exclude all particles within the central $1\;R_d$ when calculating the $m=2$ Fourier amplitudes, obtaining $A_{2, \text{excl}}$ and $\Omega_{p, \text{excl}}$ (shown in the third row of \autoref{fig:PS_A2}). 
The resulting \PSA\ distributions closely match those derived from disk-only particles, deviating only slightly during the final saturation phase. 
This consistency reinforces the conclusion that the disk component governs pattern speed evolution in the secular stage.

The bottom three rows of \autoref{fig:PS_A2} display the \PSA\ distributions for the warm model series. 
Despite the fact that most bars in the warm models do not have a secular evolution phase, their behavior in the \PSA\ space is qualitatively similar to that of the cold series.

\subsection{\LzA \ space}
\label{subsec:lza}

\begin{figure*}
  \centering
  \includegraphics[width=\textwidth]{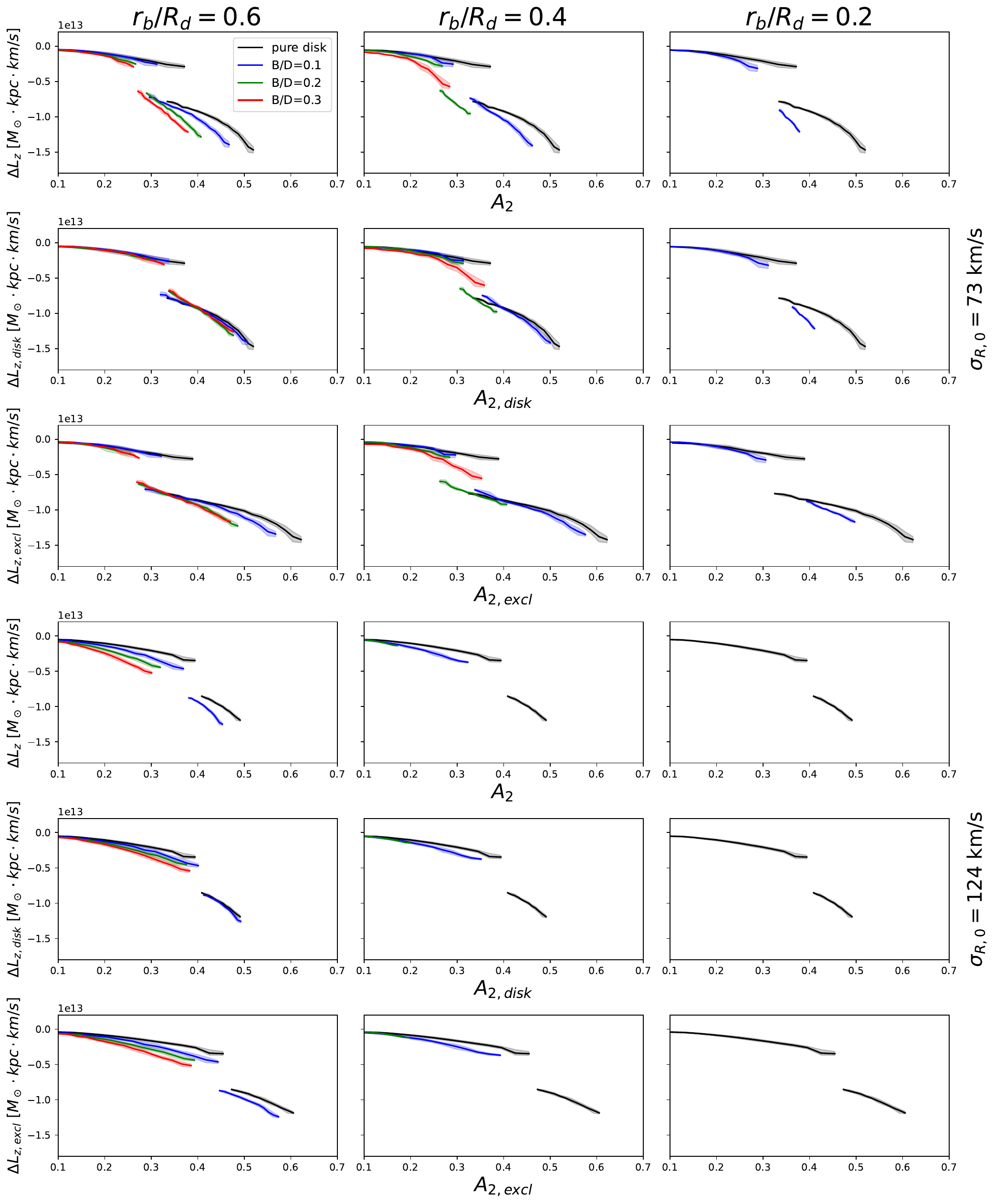}
  % \caption{test}
  \caption{Similar to \autoref{fig:PS_A2}, but in the \LzA\ space. The $y$-axis shows the change in angular momentum ($\Delta L_z$) of the stellar disk within 4\;\Rd.
  During bar formation, stronger bulges lead to greater angular momentum loss in the disk at a given bar strength. In the secular growth stage, however, models exhibit comparable $\Delta L_z$ at fixed $A_2$ once the bulge's contribution is removed or reduced.
  }
\label{fig:Lz_A2}
\end{figure*}

Our earlier analysis attributed the similar pattern speeds at fixed $A_2$ of bars formed under different mechanisms to the consistent angular momentum loss in the inner disk, as illustrated in Figure~7 of \citetalias{Zheng2025}. 
Guided by this physical picture, we now examine the angular momentum loss--bar strength (\LzA) space.
Here, $\Delta L_z$ is defined as the angular momentum loss of the stellar components within $4R_d$, i.e., the change relative to its initial value:
\begin{equation}
  \Delta L_z = L_z (<4\;R_d) - L_{z,ini}(<4\;R_d).
\end{equation}
Using the same reduction procedure as in the \PSA\ analysis, we plot the distribution of bars in the \LzA\ space.

The first row of \autoref{fig:Lz_A2} shows that all cold-series models lose the same amount of angular momentum at the onset of bar formation. 
However, models with more massive or compact bulges start with greater angular momentum, as indicated by their higher circular speeds (\autoref{fig:model}). 
Consequently, these models retain a higher residual angular momentum after this initial loss, which translates into a higher pattern speed in the early bar formation stage.
These models then exhibit a faster loss of angular momentum as their bar grows stronger, explaining the faster deceleration in pattern speed at fixed bar strength seen in the \PSA\ space.
This enhanced loss persists throughout the secular stage, resulting in a lower final pattern speed for bars in models with more massive/compact bulges. 
The consistency between the \PSA\ and \LzA\ spaces supports our proposal that the difference in pattern speed at fixed bar strength stems from differing amounts of angular momentum loss of stars.

Analogous to the \PSA\ space analysis, we also plot the \LzA\ space using disk-only measurements (second row of \autoref{fig:Lz_A2}) and measurements that exclude the central $1\;R_d$ (third row of \autoref{fig:Lz_A2}) to mitigate the bulge's influence. For these cases, the angular momentum loss $\Delta L_z$ is computed using only disk particles or by excluding particles within the central $1\;R_d$, respectively.

The \PSA\ and \LzA\ spaces present a consistent picture: once bulge contributions are minimized, models exhibit similar angular momentum loss $\Delta L_z$ at fixed $A_2$ during secular evolution, just as they show similar pattern speeds $\Omega_p$.
As a further check, we examined hybrid spaces where $\Delta L_z$ includes the bulge but $A_2$ is measured from the disk only ($A_{2,\mathrm{disk}}$) or with the center excluded ($A_{2,\mathrm{excl}}$). The key trends in these hybrid spaces are identical to those in the fully bulge-reduced analysis.

The results for the warm model series (bottom three rows of
\autoref{fig:Lz_A2}) further reinforce the consistency between
the \PSA\ and \LzA\ spaces.

Our results provide a kinematic insight for understanding the bulge's influence on bar pattern speed.  
The bulge increases the initial angular momentum of the inner stellar disk, leading to a higher pattern speed during the bar formation stage.
It also promotes a greater loss of angular momentum at a fixed growth of bar strength. 
Although the bulge itself gains angular momentum during its co-evolution with the bar, we find that the amount gained by the bulge is significantly smaller than the amount lost by the inner stellar disk, consistent with previous studies \citep{Saha2016, Kataria2018}.
We suggest that the presence of bulge components enhances the transfer of angular momentum from the inner stellar disk to the outer disk and halo within the same growth in bar strength, resulting in faster pattern speed deceleration rate. 
During the secular growth stage, the disk component dominates the evolution of both angular momentum and pattern speed, whereas the bulge's influence may become less important.

\subsection{\Rratio\ ratio}
\label{subsec:rratio}

\begin{figure*}
  \centering
  \includegraphics[width=\textwidth]{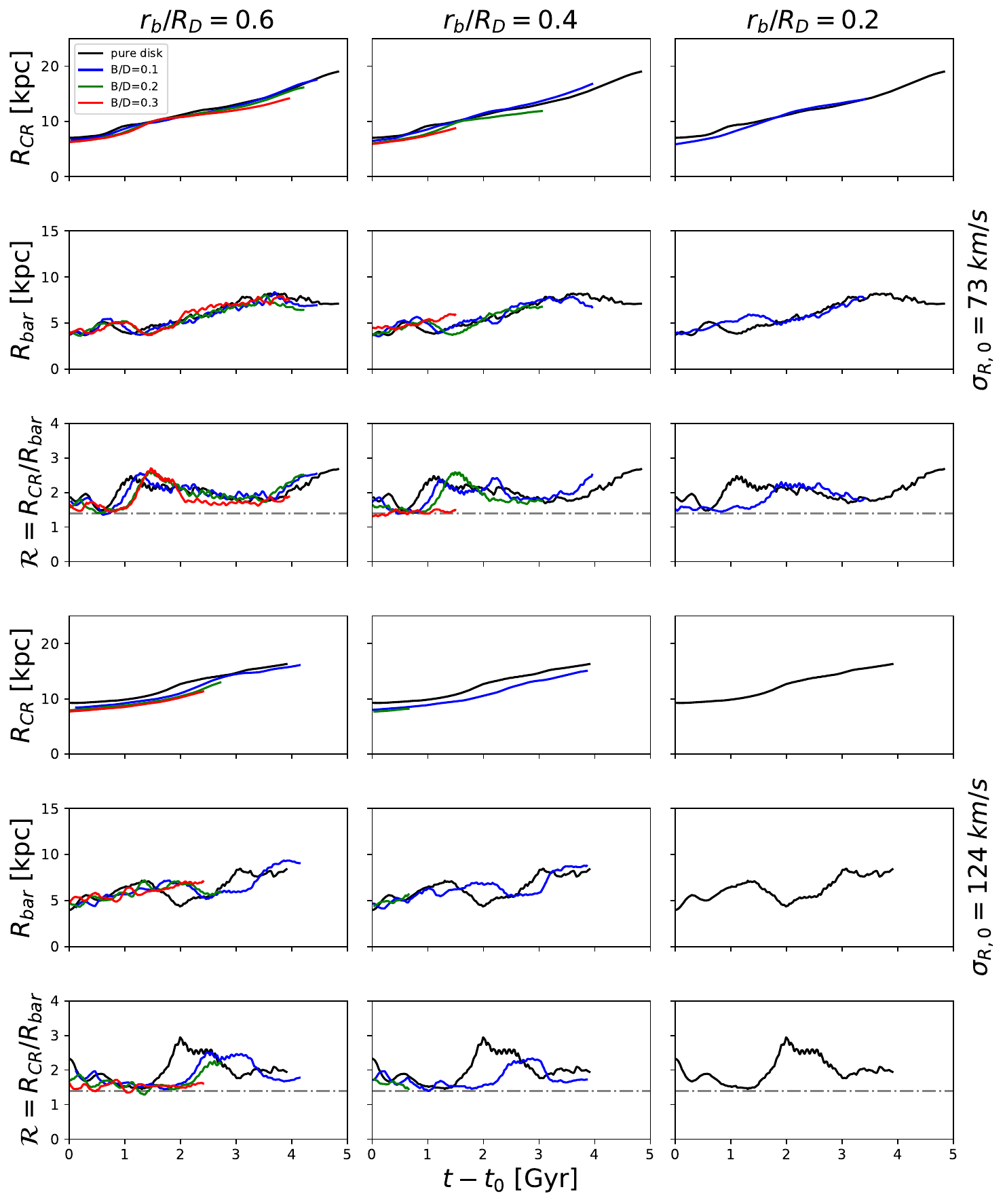}
  \caption{
  Time evolution of the co-rotation to bar-length ratio (denoted \Rratio) since bar formation ($t_0$, defined at \BS$=0.1$).
  Layout and coloring match \autoref{fig:bar_pro}. The vertical axis displays $R_{\mathrm{CR}}$, $R_{\mathrm{bar}}$, and their ratio $\mathcal{R}$. 
  The horizontal dash-dotted line indicates $\mathcal{R}=1.4$, separating fast and slow bars. 
  Reliable measurement of $R_{\mathrm{CR}}$ becomes impossible beyond ${\sim}10\;R_d$ due to low particle counts; these epochs are omitted. 
  Bars in models with a more massive bulge show a lower \Rratio\ ratio at their birth but hold a similar \Rratio\ ratio after buckling.
  }
\label{fig:Rratio}
\end{figure*}

To further investigate how the bulge affects bar pattern speed, we analyze the co-rotation to bar-length ratio, defined as $\mathcal{R} = R_{\mathrm{CR}} / R_{\mathrm{bar}}$ \citep{Debattista1998, Debattista2000}. 
Bars with $\mathcal{R} < 1.4$ are considered ``fast'', and those with $\mathcal{R} > 1.4$ are ``slow''. 
Measurements of $R_{\mathrm{CR}}$ and $R_{\mathrm{bar}}$ follow the methods detailed in \citetalias{Zheng2025} (Section~3.4 and Appendix).
Briefly, $R_{\mathrm{CR}}$ is the radius where the bar pattern speed equals the local circular frequency ($\Omega_p = \Omega_c(R_{\mathrm{CR}})$). 
The bar length $R_{\mathrm{bar}}$ is identified as the radius where the $m=2$ phase angle deviates from the bar's near-constant value by $5^\circ$. 
We apply a 0.2\Gyr\ moving average to both $R_{\mathrm{CR}}$ and $R_{\mathrm{bar}}$ to reduce fluctuations.

The evolution of the co-rotation radius $R_{\mathrm{CR}}$, bar length $R_{\mathrm{bar}}$, and their ratio $\mathcal{R}$ are shown in \autoref{fig:Rratio}. 
To compare the evolution meaningfully despite varying bar formation times, we plot them against $t-t_0$, where $t_0$ is defined as the time when the bar strength \BS\ reaches 0.1.
In all simulations, the co-rotation radius $R_{\mathrm{CR}}$ gradually increases as the bar slows down. 
Regardless of the
bulge properties, bars in all simulations show comparable lengths $R_{\mathrm{bar}}$ after formation. 
This similarity suggests that the bar length is determined primarily by the disk properties—which are common to all models—rather than by the bulge. 
When the bar buckles, its length drops rapidly, producing a bump in the evolution of $\mathcal{R}$, since $R_{\mathrm{CR}}$ does not change significantly at the same time.

Comparing different simulations, bars in models with a more massive/compact bulge show a smaller \Rratio\ ratio at the beginning of bar formation, which indicates a faster rotation rate.
However, the difference in \Rratio\ ratio becomes less significant as the bars grow.
They reach a similar \Rratio\ ratio prior to the buckling stage, which is consistent with the similar pattern speed observed at the end of the formation stage in the \PSA\ space.
After the bump caused by the buckling, bars in different models show a comparable \Rratio\ ratio, which is consistent with the similar pattern speed at fixed bar strength during the secular growth stage when the bulge contribution is removed or reduced (the second/fifth and third/sixth rows of \autoref{fig:PS_A2}).
These results further confirm that the bulge's influence on the bar pattern speed is more significant during the bar formation stage, while in the secular growth stage, the disk component dominates the pattern speed evolution, and the bulge may only have a limited impact.

% Comparing across simulations, bars in models with more massive bulges exhibit a smaller $\mathcal{R}$ ratio at the onset of formation, indicating a faster initial rotation. 
% This difference becomes less significant as the bar grows, and the ratios converge to similar values prior to buckling. This convergence is consistent with the comparable pattern speeds seen at the end of the formation stage in the \PSA\ space.
% After the transient bump induced by buckling, $\mathcal{R}$ becomes similar across models, echoing the convergence of pattern speeds at fixed bar strength during secular evolution—once bulge contributions are mitigated (rows 2, 3, 5, and 6 of \autoref{fig:PS_A2}). 
% These results collectively confirm that the bulge exerts a stronger influence on pattern speed during bar formation, whereas in the secular stage, the evolution is dominated by the disk, with the influence from the bulge likely being secondary.

\section{Bulge effects on bar growth timescale}
\label{sec:growth_timescale}

\begin{figure*}
  \centering
  \includegraphics[width=\textwidth]{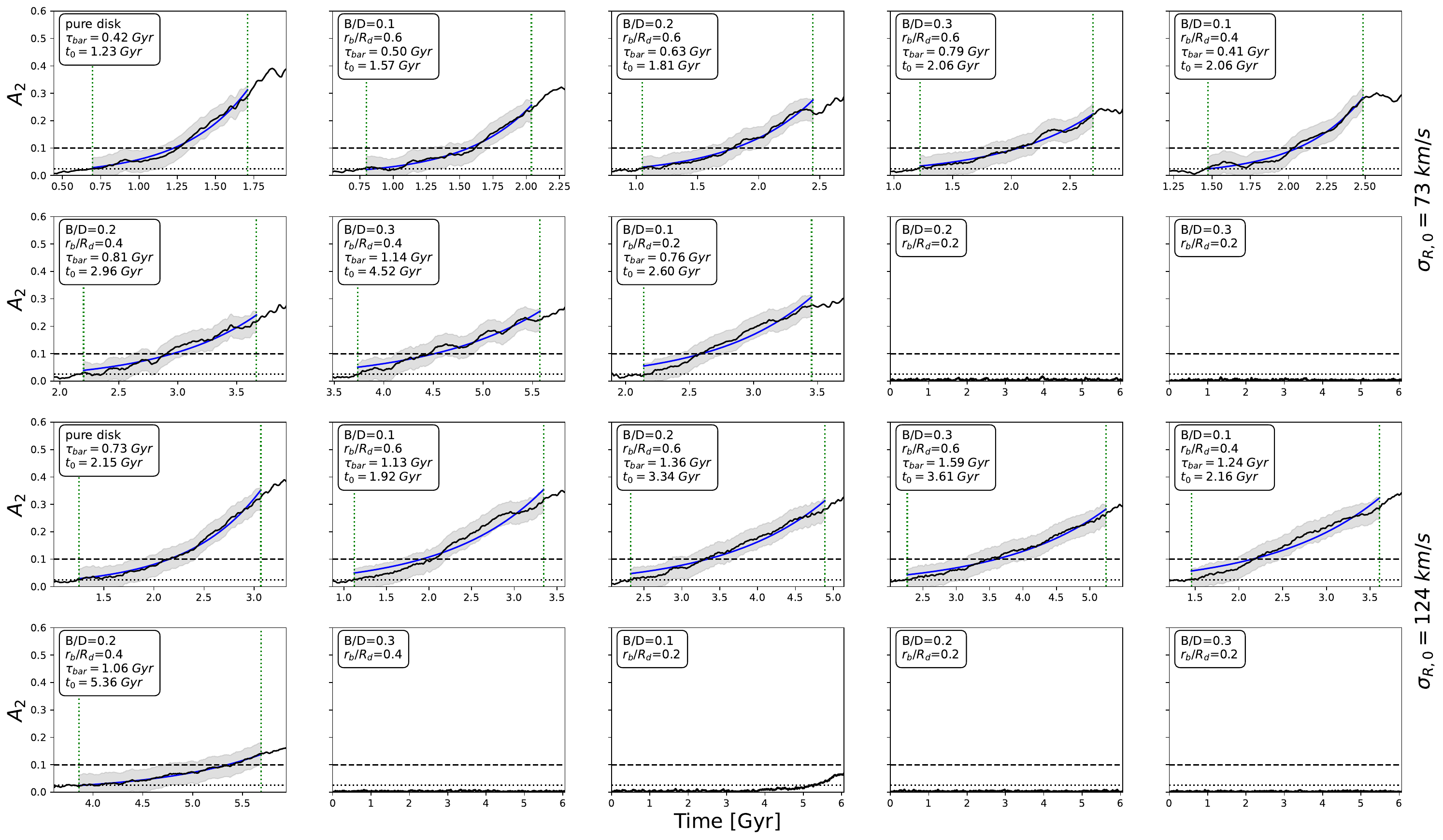}
  \caption{Exponential fitting to the $A_2$ growth.
  The evolution of $A_2$ is shown by the black solid line, with its $1\sigma$ spread indicated by the gray shading. The blue solid line represents the exponential fit. The regions used for fitting are denoted by green vertical dotted lines. 
  The text box notes the model information of bulge-to-disk mass fraction ($B/D=${0.1}, 0.2, and 0.3), the bulge scale radius to the stellar disk scale length ratio (\RadiusRatio\ $=0.6$, 0.4, and 0.2). When applicable, the fitted parameters for the bar growth timescale (\tbar) and onset time ($t_0$) are also provided.
  }
\label{fig:exp_fit}
\end{figure*}

\begin{figure*}
  \centering
  \includegraphics[width=\textwidth]{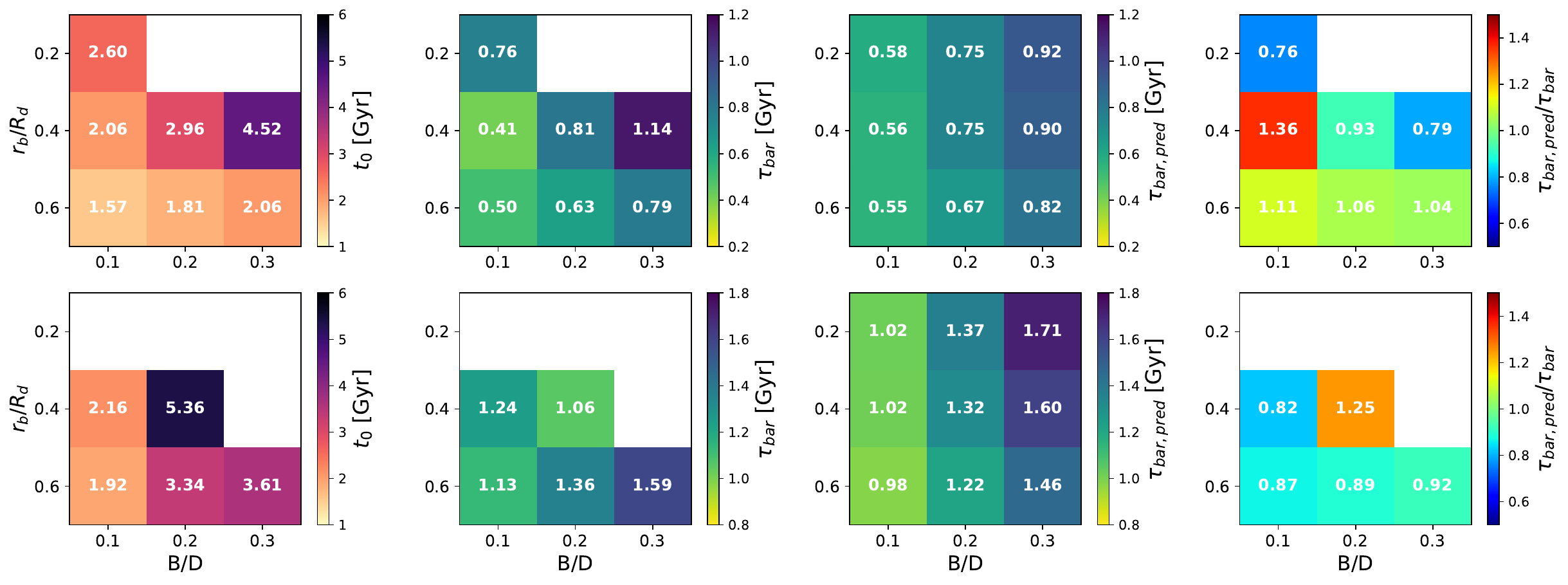}
  % \caption{test}
  \caption{
    Results of the exponential fitting for the cold series (top row) and warm series (bottom row). 
    \textbf{Left column}: fitted onset time $t_0$ (when $A_2$ reaches 0.1). More massive/compact bulges delay bar formation, yielding a larger $t_0$. 
    \textbf{Middle left column}: fitted growth timescale \tbar. More massive/compact bulges slow bar growth, leading to larger \tbar. 
    \textbf{Middle right column}: predicted growth timescale $\tau_{\mathrm{bar, pred}}$ based on \citet{Chen2025}. 
    \textbf{Right column}: ratio of the  predicted to fitted timescale, $\tau_{\mathrm{bar, pred}}/\tau_{\mathrm{bar}}$.  
  }
\label{fig:fit_info}
\end{figure*}

Following \citet{BlandHawthorn2023}, we quantify the bar growth timescale by fitting an exponential function to the $A_2$ growth during the bar formation stage.
The function is defined as:
\begin{equation}
\label{eq:exp_fit}
A_2 (t) = 0.1  \exp((t - t_0)/\tau_{\mathrm{bar}}),
\end{equation}
where $t_0$ (the time when $A_2=0.1$) marks the bar onset, and $\tau_{\mathrm{bar}}$ quantifies the growth timescale.
\citetalias{Zheng_taubar2025b} demonstrated that different exponential fitting forms produce consistent $\tau_{\mathrm{bar}}$ values with variations of approximately $20\%$ while keeping comparable goodness-of-fit. 
Consequently, we treat $20\%$ as an acceptable uncertainty for $\tau_{\mathrm{bar}}$. 
Full details of the fitting methodology and its validation are given in Section~3 and the Appendix of \citetalias{Zheng_taubar2025b}.

The fitting results are presented in \autoref{fig:exp_fit}, with the fitted parameters noted in each panel's text box where applicable.
Most simulations display an exponential growth in bar strength,
and the fitted curves capture the $A_2$ evolution accurately.
This behavior aligns with the prediction by the swing amplification feedback loop.

We then compare the fitted parameters in \autoref{fig:fit_info}.  The left column shows the fitted onset times $t_0$. 
A more massive bulge delays bar formation, yielding a larger $t_0$.  
Similarly, for a fixed bulge mass, a more compact bulge also delays the onset. 
This is consistent with the stronger stabilizing effect that a more massive or compact bulge exerts on the inner stellar disk, as reflected in the higher Toomre $Q$ profiles shown in \autoref{fig:model}.

The middle left column displays the fitted growth timescale \tbar. 
According to the Fujii relation \citep{Fujii2018, BlandHawthorn2023}, the bar growth timescale is sensitive to $f_{\text{disk}}$, the disk mass fraction. 
Since a more massive bulge reduces $f_{\text{disk}}$, it should—and indeed does—produce a longer \tbar, indicating slower bar growth.  
A more compact bulge likewise tends to increase \tbar, with a few exceptions\footnote{Specifically, the cold model with $B/D=0.1$, \RadiusRatio$=0.2$ and the warm model with $B/D=0.2$, \RadiusRatio$=0.4$.}.

Based on \textit{pure disk models}, \citet{Chen2025} proposed an empirical relation to predict the bar growth timescale using the disk mass fraction ($f_{\text{disk}}(2.2R_d)$), Toomre $Q(2.2R_d)$, and the disk scale height ($h_z$):
\begin{equation}
  \label{eq:chen_relation}
\frac{\tau_{\mathrm{bar, pred}}}{C} = Q(2.2\;R_d) \frac{h_{z}}{R_{d} } \exp \left(-\frac{f_{\text{disk}}(2.2\;R_d)}{0.110}\right),
\end{equation}
We apply this relation to our galaxies with bulges to predict $\tau_{\mathrm{bar, pred}}$ and compare it with the fitted \tbar\ values. This comparison aims to identify potential directions for improving the prediction of bar growth timescales in galaxies with bulges.

The right-hand side of \autoref{eq:chen_relation} is
dimensionless and irrelevant to the simulation scaling.
Therefore, it does not set the physical time unit; the parameter $C$ must be calibrated to match our simulation scaling.
We calibrate $C$ using the fitted \tbar\ value and the relevant parameters of the isolated pure disk model in the cold series, obtaining $C = 375$\Gyr.
\footnote{ Calibrating with the warm pure disk model yields a slightly different $C$ but does not alter our conclusions.}
With this calibrated $C$, we compute the predicted bar growth timescale for our models containing a bulge via \autoref{eq:chen_relation} and compare it with the fitted \tbar\ values (the middle-right and right columns of \autoref{fig:fit_info}).

The predicted growth timescale shows reasonable agreement with the fitted \tbar\ values (within the ${\sim}20\%$ uncertainty) for models with diffuse bulges (\RadiusRatio$=0.6$).  
Moreover, the relation correctly reproduces the trend of longer \tbar\ for more massive bulges. 
This agreement implies that the formula of \citet{Chen2025}—via its influence on $f_{\text{disk}}(2.2R_d)$—may already effectively account for the effect of bulge mass on the bar growth rate.

A key limitation of \autoref{eq:chen_relation} is its inability to account for bulge compactness: it yields similar predicted $\tau_{\mathrm{bar, pred}}$ values for models sharing the same bulge mass but different \RadiusRatio. 
This shortcoming is starkly evident in highly compact bulges that completely suppress bar formation within 6\Gyr, implying a much longer true growth timescale than predicted. 
The root cause is that both $Q(2.2R_d)$ and $f_{\text{disk}}(2.2R_d)$—the formula's inputs—are nearly identical for such models (see \autoref{fig:model}), rendering them insensitive to compactness. 
This indicates that either (1) we should measure $Q$ and $f_{\text{disk}}$ at a smaller radius, where compactness has a stronger effect, or (2) a compactness parameter should be added explicitly to the formula.

A more definitive assessment of how bulge compactness affects the bar growth timescale would require a larger suite of simulations and a more systematic analysis, which is beyond the scope of the present study and will be pursued in future work.

%-----------
%-- Sect. 5
%-----------

\section{Summary}
\label{sec:summary}

We conduct a suite of controlled $N$-body simulations to examine the effect of classical bulges on spontaneous bar
formation and subsequent bar properties. 
We generate a series of galaxy models with bulges spanning a range of bulge-to-disk mass ratios and compactness. 
A more massive or compact bulge increases the Toomre $Q$ parameter and reduces the disk mass fraction ($f_{\text{disk}}$) in the central region(\autoref{fig:model}).

We evolve these galaxy models in isolation and measure the resulting bar strength and pattern speed (\autoref{fig:bar_pro}). 
A more massive or compact bulge delays bar formation and seems to produce a higher pattern speed. 
If the bulge is sufficiently massive and compact, bar formation is completely suppressed within the 6\Gyr\ simulation timeframe.

To facilitate a fair comparison of bar pattern speeds, we construct the pattern speed--bar strength (\PSA) space, which accounts for differences in evolutionary stage (\autoref{fig:PS_A2}). 
In the early formation stage, bars in models with more massive or compact bulges have a higher initial pattern speed \PS\ but decelerate more rapidly as the bar strengthens. 
This accelerated deceleration persists into the secular growth stage, leading to a lower \PS\ at a fixed bar strength. 
After removing or reducing the bulge's ``diluting'' effect on the measured bar strength, the same trends—higher initial \PS\ and faster deceleration—persist in the bar formation stage.
However, the pattern speeds become similar in the secular stage. 
These results indicate that the bulge strongly influences the pattern speed during bar formation, but its role may become less important in the secular stage, where the disk component dominates the evolution of \PS.

A kinematic insight of these results is provided in the
angular momentum loss--bar strength (\LzA) space (\autoref{fig:Lz_A2}).
A more massive or compact bulge increases the initial angular momentum of the inner disk, leading to a higher initial pattern speed. 
The bulge promotes the transfer of angular momentum from the inner stellar disk to the outer regions and the halo, driving a faster deceleration in pattern speed.
During the secular growth stage, the disk component dominates the angular momentum evolution and consequently the evolution of the pattern speed.

The trend in pattern speed evolution is also reflected by the \Rratio\ ratio (\autoref{fig:Rratio}).
Bars in models with more massive/compact bulges have a smaller \Rratio\ at the early stage of bar formation, but the difference in \Rratio\ becomes smaller during the growth of the bar and eventually becomes indistinguishable in the secular growth stage.
These results echo the conclusion from the \PSA\ space that the bulge's effect on the pattern speed is more significant during the formation stage but can be less important in the secular growth stage.

% The evolution of the \Rratio\ parameter (\autoref{fig:Rratio}) provides additional, independent support for our main finding.
% It shows that bars in models with more massive/compact bulges initially have a smaller \Rratio\ (indicating faster rotation), but this difference decreases as the bar strengthens and disappears in the secular growth stage. 
% The consistency between the \Rratio\ evolution and the behavior in the \PSA\ space underscores the same picture: the bulge's impact on bar pattern speed is primary during the formation stage, but may become secondary during the secular growth stage.

We quantify the bar growth by applying an exponential fit to the $A_2$ evolution to measure the onset time $t_0$ and the growth timescale \tbar\ (\autoref{fig:exp_fit}). 
More massive or compact bulges delay bar formation (larger
$t_0$), and the bars in these models grow more slowly, as
reflected in their larger \tbar.

This paper investigates the effect of classical bulges on bar formation and properties under the internal formation mechanism. 
To build a more complete picture, the subsequent study (Paper IIIB) will explore the role of bulges in tidally induced bar formation. The combined analysis of both internal and external formation scenarios will allow for a direct comparison of how bulge properties influence bars under these distinct mechanisms.

%%%%%%%%%%%%%%%%%%%%%%%%%%%%%%%%%%%%%%%%%%%%%%%

\software{
    {\sc agama}\citep{AGAMA2019},
    \texttt{GADGET-4} \citep{Springel2005,Springel2021}. 
    NumPy \citep{2020NumPy-Array},
    SciPy \citep{2020SciPy-NMeth},
    Matplotlib \citep{4160265},
    Jupyter Notebook \citep{Kluyver2016jupyter}
}

% \begin{acknowledgments}
\section*{Acknowledgements}
% We thank the referee for suggestions that helped to improve the presentation of the paper. 
We thank Thor Tepper-Garcia and Joss Bland-Hawthorn for their help with the \agama\ work and with initializing the galaxy models.
We thank Xufen Wu, Zhi Li, and Sandeep Kumar Kataria for their valuable insights on simulations and analysis. We also thank Rui Guo for their helpful discussions.
% FUNDING!!!
The research presented here is partially supported by the National Natural Science Foundation of China under grant Nos. 12533004, 12025302, 11773052; by China Manned Space Program with grant no. CMS-CSST-2025-A11; and by the “111” Project of the Ministry of Education of China under grant No. B20019. 
% The research presented here is partially supported by the National Key R\&D Program of China under grant No. 2018YFA0404501; by the National Natural Science Foundation of China under grant Nos.  12025302, 11773052, and 11761131016; by the ``111'' Project of the Ministry of Education of China under grant No. B20019; and by China Manned Space Program with grant No. CMS-CSST-2025-A11. J.S. acknowledges support from the {\it Newton Advanced Fellowship} awarded by the Royal Society and the Newton Fund. 
% X.W. wishes to thank the financial support from the Natural Science Foundation of China (Numbers NSFC-12073026 and NSFC-12433002).
This work made use of the Gravity Supercomputer at the Department of Astronomy, Shanghai Jiao Tong University.
% \end{acknowledgments}

%%%%%%%%%%%%%%%%%%%%%%%%%%%%%%%%%%%%%%%%%%%%%%%%%%

%%%%%%%%%%%%%%%%% APPENDICES %%%%%%%%%%%%%%%%%%%%%

% \appendix
% \restartappendixnumbering
% \section{Fitting validation}
% \label{app:diff_fit}

% % different fitting formulas yield similar $\tau_{\mathrm{bar}}$--> appendix, figure
% In addition to \autoref{eq:exp_fit}, we explored alternative exponential formulas to fit the $A_2$ evolution. One such formula is

% \newpage

\bibliography{tidal_bar.bib}
\bibliographystyle{aasjournal}

\end{document}